\documentclass[%
reprint,
superscriptaddress,
amsmath,amssymb,
aps,
pre,
]{revtex4-1}

\usepackage{graphicx}% Include figure files
\usepackage{mathtools} % Adds on amsmath
\usepackage{siunitx} % For units
\usepackage{bbold} % For \mathbb{1} etc., the standard doesn't work with numbers

\newcommand*\diff{\mathop{}\!\mathrm{d}}
\newcommand{\Mod}[1]{\ (\mathrm{mod}\ #1)}
\newcommand{\overbar}[1]{\mkern 1.5mu\overline{\mkern-1.5mu#1\mkern-1.5mu}\mkern 1.5mu}

\DeclareMathOperator{\arccosh}{arccosh} % Print arccosh as an operator

\usepackage{dcolumn}% Align table columns on decimal point
\usepackage{bm}% bold math
\usepackage{afterpage}

\begin{document}
	
	%\preprint{APS/123-QED}
	
	\title{Diffusible crosslinkers cause superexponential friction forces}
	% Force line breaks with \\
	%\thanks{}%
	
	\author{Harmen Wierenga}
	\affiliation{AMOLF, Science Park 104, 1098 XG Amsterdam, The Netherlands}
	\author{Pieter Rein ten Wolde}
	\email{p.t.wolde@amolf.nl}
	\affiliation{AMOLF, Science Park 104, 1098 XG Amsterdam, The Netherlands}

	
	\begin{abstract}
		The mitotic spindle lies at the heart of the spatio-temporal control over cellular components during cell division. 
		The spindle consists of microtubules, which are not only crosslinked by motor proteins 
		but also by passive binding proteins.
		These passive crosslinkers stabilize the highly dynamic mitotic spindle 
		by generating friction forces between sliding filaments.
		However, it remains unclear how the friction coefficient depends on 
		the number of crosslinkers and the size of the overlap between the microtubules.
		Here, we use theory and computer simulations to study 
		the friction between two filaments that are crosslinked by passive proteins,
		which can hop between neighboring binding sites while physically excluding each other.
		The simulations reveal that
		the movement of one microtubule relative to the other is limited by free-energy barrier crossings, 
		causing rare and discrete jumps of the microtubule that span the distance between adjacent crosslinker binding sites.
		We derive an exact analytical expression for the free-energy landscape 
		and identify the reaction coordinate that governs the relative movement, 
		which allows us to determine the effective barrier height for the microtubule jumps.
		Both through simulations and reaction rate theory,
		we make the experimentally testable prediction that 
		the friction between the microtubules increases superexponentially with the density of crosslinkers.

		%\begin{description}
		%use the optional argument of the \verb+\item+ command to give the category of 
		%each %item. 
		%\end{description}
	\end{abstract}
	
	\pacs{Valid PACS appear here}% PACS, the Physics and Astronomy
	\maketitle
	
	%\tableofcontents

\section{\label{sec:introduction} Introduction} 
A key structure in eukaryotic cell division is the mitotic spindle~\cite{inoue_cell_1981},
which controls chromosome segregation and cytokinesis~\cite{alberts_molecular_1994}.
The mitotic spindle consists of microtubules~\cite{alberts_molecular_1994}
that grow radially outwards from two opposing poles before cell division~\cite{karsenti_mitotic_2001}.
The geometry of the spindle is crucial for its function:
the proper establishment of the metaphase plate,
the region where growing microtubules meet and overlap each other in the cell midzone,
is essential for division into equally sized daughter cells~\cite{tan_equatorial_2015}.
Yet, how a stable spindle is organized remains poorly understood.

The spindle structure is formed by a dynamic interplay between active driving forces generated by motor proteins
and stabilizing forces that counteract these motor forces.
Plus-end directed motor proteins, such as kinesin-5,
crosslink the antiparallel microtubules that come from the opposing sides,
and attempt to slide them apart~\cite{kapitein_bipolar_2005}.
This activity must be counterbalanced, because otherwise
the motor forces would prevent the interdigitation of the microtubules in the midzone~\cite{mollinari_prc1_2002}
and disrupt cytokinesis~\cite{jiang_prc1:_1998}.
Intriguingly, antagonistic minus-ended directed motor proteins,
which oppose the pulling forces of the plus-end directed motors,
are not sufficient because the combination of plus- and minus-end directed motors alone
yields unstable structures~\cite{hentrich_microtubule_2010}.
To create a stable spindle, the motor forces must be balanced via stabilizing forces~\cite{hentrich_microtubule_2010}.

Friction is likely to be a key stabilizing force~\cite{forth_mechanics_2017}.
While the friction forces between sliding actin filaments can be significant,
the friction between bare, unlinked microtubules is negligible~\cite{ward_solid_2015}.
Motor proteins do not only generate an active driving force, but also
a friction force when they are pulled faster than their intrinsic motor velocity~\cite{shimamoto_measuring_2015}.
However, the observation that motor proteins alone cannot form a stable overlap~\cite{hentrich_microtubule_2010}
indicates that this friction is not sufficient.

Interestingly, the mitotic spindle also contains proteins that crosslink the microtubules passively,
and the friction forces generated by these passive crosslinkers
are likely to be essential for the formation of a stable spindle~\cite{hentrich_microtubule_2010,forth_asymmetric_2014}.
Examples of such non-motor proteins are 
Ase1 in yeast~\cite{schuyler_molecular_2003} and PRC1 in mammals~\cite{jiang_prc1:_1998}.
Previously, it has been shown that Ase1 has a strong affinity for the overlap region between two microtubules, 
and that the proteins are able to diffuse within the microtubule overlap~\cite{kapitein_microtubule-driven_2008}. 
This thermal motion of the crosslinkers creates entropic forces that can antagonize motor forces 
and lead to a steady state overlap size 
when the microtubules are not dynamic~\cite{lansky_diffusible_2015,johann_assembly_2016}.
However, many of the short spindle microtubules have lifetimes of roughly $\SI{20}{\second}$~\cite{oriola_physics_2015},
which may prevent microtubule overlaps from reaching the steady-state size 
that is determined by the balance between the motor forces 
and the entropic force generated by the passive crosslinkers.
Additionally, microtubules constantly slide past each other due to polymer growth and motor activity,
causing a net polewards flux of microtubules~\cite{mitchison_polewards_1989}.
Crosslinking proteins generate friction forces between these sliding microtubules,
which can play a large role in the force balance 
in steady state spindles~\cite{hentrich_microtubule_2010,forth_asymmetric_2014},
specifically in the antiparallel overlap region located at the midzone where PRC1 binds~\cite{jiang_prc1:_1998}.
To understand the size of the spindle structure and the timescale on which it is formed, 
we need to know how the friction between sliding microtubules depends on 
the size of the overlap region and the number of crosslinkers bound to it~\cite{forth_mechanics_2017}.

To study crosslinker generated friction forces, 
we analyze the model that was previously used to study entropic force generation~\cite{lansky_diffusible_2015}.
There, it was found computationally and confirmed experimentally
that the friction coefficient between two microtubules increases exponentially with the number of crosslinkers.
Nevertheless, the origin of this behavior remains elusive,
and a theoretical description of the friction coefficient is lacking.

Here, we provide a theoretical characterization of the friction coefficient. 
We show that relative movement of two microtubules is governed by 
rare jumps spanning exactly one tubulin dimer length.
The jumps are the result of the discrete hopping of crosslinkers between binding sites
on the different microtubule subunits,
and the rare events can be effectively described as free-energy barrier crossings.
Using this framework, we find an analytical solution for the friction coefficient. 
In the limit of low crosslinker densities 
we retrieve the exponential behavior that was observed experimentally~\cite{lansky_diffusible_2015}, 
and we obtain an expression for the exponent. 
Surprisingly, however, at higher densities the friction increases superexponentially with the crosslinker density.

We hypothesize that the superexponential friction dependence is utilized by cells 
to effectively stall sliding microtubules at well defined overlap lengths in the mitotic spindle.
Additionally, the predicted friction dependence can be used to contrast different models for crosslinker binding.
Models that allow the crosslinkers to diffuse smoothly over the microtubule lattice 
predict a linear crosslinker dependence for the friction coefficient,
whereas the highly non-linear friction observed here relies on the discrete nature of the binding sites.

\section{\label{sec:model} Model} 
\begin{figure}
	\includegraphics{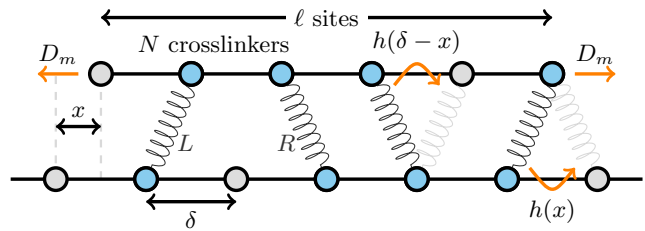}
	\caption{\label{fig:model}
		Model of the microtubule overlap. 
		The filaments are represented by one dimensional lattices, 
		that align and are connected by crosslinking proteins,
		represented by springs. 
		The bottom filament is fixed, while the top filament can move its longitudinal direction
		due to Brownian motion with diffusion constant $D_{m}$ 
		or the pulling of stretched crosslinkers. 
		When the mobile filament moves a distance $x$ to the right, 
		all crosslinkers that were previously straight are now stretched, 
		pulling the filament back to the left.
		These left-pulling linkers are labeled by $L$, whereas right-pulling linkers are labeled by $R$. 
		It is assumed that no crosslinker stretches further than one lattice spacing $\delta$.
		Crosslinkers can make a diffusive step with one head, 
		causing them to switch from $R$ to $L$, or vice versa.
		Two of such possible transitions, together with their rates $h$, are denoted by orange arrows.
		We assume the fixed microtubule to be infinitely long and the mobile microtubule to have $\ell$ lattice sites.
		There are $N$ crosslinkers connecting the two filaments, which stay bound indefinitely.
		}
\end{figure}
Our goal is to understand how the friction between two microtubules depends on
the absolute number of proteins crosslinking them and on the length of their overlap region.
For this, we turn to the model shown in Fig.\,\ref{fig:model}
that captures the key mechanisms by which crosslinkers influence microtubule movement.
This model was previously successful in explaining the entropic force generation by diffusive crosslinkers,
and it resembles the experimental setup used to study the properties of Ase1 proteins 
in vitro~\cite{lansky_diffusible_2015}.
We imagine a microtubule that is fixed on the bottom, 
and a second microtubule on top of the first one connected via crosslinkers.
The top microtubule can move in one dimension, parallel to the fixed microtubule.
Since we are interested in the friction between these two filaments, 
we avoid entropic force generation by ensuring that the overlap length remains constant~\cite{lansky_diffusible_2015}.
To this end, the fixed filament is assumed to be much longer than the mobile one, 
such that the two microtubules always fully overlap, as indicated in Fig.\,\ref{fig:model}.

The microtubules consist of tubulin dimers, 
which each contain one binding site for the crosslinking proteins~\cite{kellogg_near-atomic_2016}. 
These binding sites are spaced at a lattice spacing of $\delta=\SI{8}{\nano\metre}$~\cite{amos_arrangement_1974}.
While each microtubule consists of $13$ protofilaments,
we assume that the crosslinkers can only bind to the one protofilament on each microtubule that is facing the other.
Therefore, the microtubules can be modeled as one dimensional lattices onto which the crosslinkers can bind,
as shown in Fig.\,\ref{fig:model}.
Crosslinkers bind between the microtubules and physically exclude each other, 
meaning that they can neither cross each other nor bind to the same site.
We also assume that the two microtubules are separated by a fixed distance.
Then, to allow for movement of both the microtubule and the linkers,
we suppose that the crosslinkers can stretch.

To enable the stretching of crosslinkers, we model them as Hookean springs.
The relaxed conformation is a straight spring between two filaments that have their lattice sites in register,
and linkers resist deformations from this energy minimum with a harmonic potential with spring constant $k$. 
The quadratic potential is a second order approximation to the full potential,
and it captures not only contributions from the physical extension of the linkers, 
but also from deviations from the preferred angle of binding.
The parameter $k$ is estimated from experimental data by observing how the diffusion constant of crosslinkers
is reduced in overlaps compared to protein diffusion on a single microtubule,
as described in detail in supplemental Sec.\,\ref{sec:estimate_spring_constant}.
This gives us the relatively high value of $k=\SI{1.1e5}{k_{B}T\per\micro\metre\squared}$, 
which makes it difficult for the microtubule to stretch the crosslinkers far.
To facilitate model analysis and speed up simulations, we choose to impose a maximum stretch on the linkers 
at a distance which is already rendered unlikely by the harmonic potential.
Specifically, we let the linkers extend at most one lattice spacing, 
at which point a single spring has a potential energy of $\SI{3.5}{k_{B}T}$.

Pulling forces from the springs cause the top microtubule to move parallel to its orientation.
The position of the mobile microtubule relative to the fixed one is called $x$.
Hence, the two microtubule lattices are perfectly aligned when ${x \equiv 0 \Mod{\delta}}$.
Unless indicated otherwise, we will intend $x$ to represent the position modulo division by $\delta$, 
such that it represents the misalignment of the two filament latices.
In that definition, the requirement that the crosslinking springs are extended less than one lattice spacing 
only allows for springs that are extended horizontally by a distance of either $x$ or $\delta-x$. 
As illustrated in Fig.\,\ref{fig:model},
these are respectively called left- and right-pulling crosslinkers,
since they apply a left- or rightward force on the mobile top-microtubule.
The microtubule responds to the net force, and we model its movement using Brownian dynamics,
as described in supplemental Sec.\,\ref{sec:simulation_dynamics}.

Besides the top microtubule, also the crosslinkers are dynamic.
To model the Brownian motion of crosslinkers within the overlap, 
both ends of the linkers can hop to neighboring sites.
This thermally driven process needs to obey detailed balance,
\begin{equation}
\frac{h\!\left(x\right)}{h\!\left(\delta-x\right)} = \exp\left(-\beta \Delta U\!\left(x\right)\right),
\label{eq:crosslinker_hopping_rate}
\end{equation}
where $h\!\left(x\right)$ is the rate at which one head of a left-pulling linker hops to a right-pulling position,
and $\beta = 1/k_{B}T$.
The system is invariant under reflections where $R\leftrightarrow L$ and $x\leftrightarrow\delta-x$, 
so the reverse process where the head of a right-pulling linker hops occurs at the rate $h\!\left(\delta-x\right)$. 
The potential energy difference $\Delta U\!\left(x\right)$ between a left- and right-pulling linker is given by
\begin{equation}
\Delta U\!\left(x\right) = U_{R}\!\left(x\right) - U_{L}\!\left(x\right)
= \frac{1}{2} k \left(\delta-x\right)^{2} - \frac{1}{2} k x^{2}.
\label{eq:potential_energy_difference_R_L}
\end{equation}
We choose the simplest rate function $h\!\left(x\right)$ that obeys detailed balance and the symmetry of the system,
\begin{equation}
h\!\left(x\right) = h_{0} \exp\left(-\frac{1}{2}\beta \Delta U\!\left(x\right)\right).
\label{eq:rate_h_form}
\end{equation}
Here, $h_{0}$ is the hopping rate when there is no change in stretch upon a hop, i.e. for $x=\delta/2$.
To estimate $h_{0}$, we also assume that a crosslinker diffusing on a single microtubule hops with this rate.
In the simulations, we implement the hops through a kinetic Monte Carlo algorithm~\cite{prados_dynamical_1997},
which we describe in more detail in supplemental Sec.\,\ref{sec:simulation_dynamics}.

Ase1 (un)binding from the microtubule overlap plays no role in 
the in vitro friction experiments~\cite{lansky_diffusible_2015},
leading us to exclude these reactions.
Ignoring binding effects reduces the number of parameters, 
and allows us to focus on the specific dependence of friction 
on the absolute number of crosslinkers in the microtubule overlap region $N$,
and on the number of lattice sites on the mobile microtubule $\ell$.

\section{\label{sec:movement} Barrier crossings cause microtubule jumps}
\begin{figure}
\includegraphics{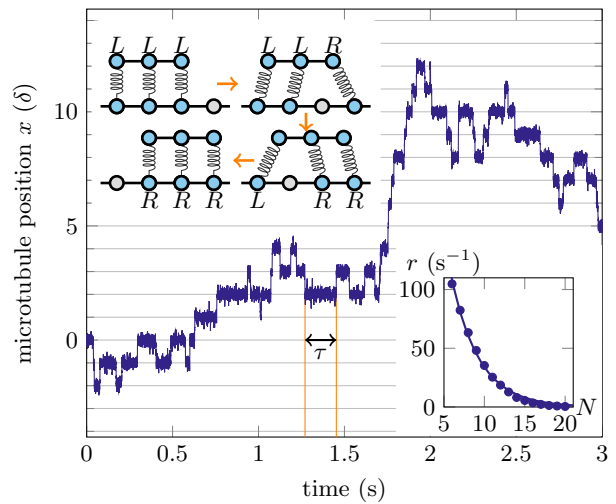}
\caption{\label{fig:position_vs_time_inset_transition_and_rate_vs_N}
	A typical time trace of the mobile microtubule position shows that it moves with sudden jumps.
	Horizontal lines denote positions where	the microtubules are aligned,
	which are $\SI{8}{\nano\metre}$ apart.
	The springs are energetically relaxed at these positions, and intermediate locations are rarely visited.
	The jumps occur at a fixed rate, which can be estimated from simulations 
	as the inverse of the mean waiting time, $r=1/\overbar{\tau}$.
	(inset top-left) A typical transition, where the microtubules begin and end in aligned positions.
	Crosslinkers are stretched in intermediate states, which energetically suppresses transitions.
	(inset bottom-right) The observed rate of microtubule jumps appears to decrease 
	exponentially with the number of crosslinkers $N$. 
	Dots show simulation estimates of the rate, 
	whereas the line shows a least square exponential fit to guide the eye. 
	In the examples, $\ell=40$, and $N=12$ for the time trace.}
\end{figure}
We are interested in the dynamics of the mobile microtubule, which we visualize using computer simulations 
and show in Fig.\,\ref{fig:position_vs_time_inset_transition_and_rate_vs_N}.
We see that the position of the top filament makes discrete jumps of exactly one lattice spacing.
Crosslinkers are energetically relaxed when their binding sites are positioned exactly above each other,
and thus microtubule positions that minimize the spring energies are those where 
the microtubules are aligned and $x\equiv0\Mod{\delta}$.
These preferential positions are indicated by horizontal lines in
Fig.\,\ref{fig:position_vs_time_inset_transition_and_rate_vs_N},
and barriers must exist between these positions to explain the rarity of the jumps.
The waiting time between jumps is exponentially distributed and has a rate $r$, 
as shown in supplemental Sec.\,\ref{sec:microtubule_rate}.

The jumping behavior causes effective Brownian motion of the microtubule with diffusion constant $D=\delta^{2} r$.
We can estimate the effective friction coefficient of the crosslinked microtubule $\zeta$ 
using the Einstein relation~\cite{einstein_uber_1905},
\begin{equation}
\zeta = \frac{k_{B}T}{D} = \frac{k_{B}T}{\delta^{2} r}.
\label{eq:einstein_relation}
\end{equation}
Previously, the friction force of a single PRC1 protein bound on a single microtubule was measured directly, 
showing a linear force-velocity relationship 
up to roughly $\SI{4}{\micro\metre\per\second}$~\cite{forth_asymmetric_2014}.
Furthermore, the diffusion constant and friction coefficient 
that were obtained in two separate experiments in~\cite{forth_asymmetric_2014}
can be compared directly to show that the Einstein relation holds for this protein,
as is the case for motors in equilibrium~\cite{bormuth_protein_2009}.
Hence, we can focus our attention on the jump rare $r$,
which indirectly gives the friction coefficient via Eq.\,\ref{eq:einstein_relation}.

The bottom-right inset of Fig.\,\ref{fig:position_vs_time_inset_transition_and_rate_vs_N} 
shows that the jump rate decreases roughly exponentially with the number of crosslinkers in the overlap $N$.
This suggests that the friction coefficient in Eq.\,\ref{eq:einstein_relation}
increases exponentially with the number of crosslinkers,
whereas one would naively expect friction to increase linearly with $N$.

To investigate the origin of the exponential decrease of the jump rate,
we calculate the free-energy landscape as a function of two order parameters involved in the jumps.
Without loss of generality, we focus on a jump to the right.
As shown in the top-left inset of Fig.\,\ref{fig:position_vs_time_inset_transition_and_rate_vs_N},
a jump requires the microtubule to move one lattice spacing,
and all crosslinkers need to make one net hop.
The former change is captured by the microtubule position $x$ changing from $0$ to $\delta$,
and the latter change is described by the number of right-pulling crosslinkers $N_{R}$ changing from $0$ to $N$.
To find the free-energy as a function of the order parameters $x$ and $N_{R}$,
we first calculate the potential energy of the system,
\begin{align}
U\!&\left(x,N_{R}\right) = \frac{1}{2} k x^{2} \left(N-N_{R}\right) 
+ \frac{1}{2} k \left(\delta-x\right)^{2} N_{R} \nonumber \\*
&= \frac{1}{2} k \delta^{2} N \left[ \left(\frac{x}{\delta} - \frac{N_{R}}{N}\right)^{2}
+ \frac{N_{R}}{N}\left(1-\frac{N_{R}}{N}\right)\right].
\label{eq:potential_energy}
\end{align}
All potential energy is stored in the springs, and there are $N_{R}$ right-pulling linkers with stretch $\delta-x$
and $N-N_{R}$ left-pulling linkers with stretch $x$.
When the values of the order parameters $x$ and $N_{R}$ are set, all microstates have the same potential energy.
Hence, we can make use of Boltzmann's formula $S=k_{B}\log \Omega$ to calculate the entropy of the system,
where, $\Omega\!\left(x,N_{R}\right)$ represents the number of microstates 
due to different permutations of the crosslinkers in the overlap.
Furthermore, the number of different permutations of the $L$- and $R$-linkers is independent of the position $x$,
meaning that $\Omega\!\left(x,N_{R}\right)=\Omega\!\left(N_{R}\right)$.
The Helmholtz free-energy is thus
\begin{align}
\mathcal{F}\!\left(x,N_{R}\right) &= U\!\left(x,N_{R}\right) - T S\!\left(x,N_{R}\right) \nonumber \\*
&=  U\!\left(x,N_{R}\right) - k_{B} T \log \Omega\!\left(N_{R}\right).
\label{eq:helmholtz_free_energy}
\end{align}
We managed to find a closed form expression for $\Omega\!\left(N_{R}\right)$,
of which the derivation can be found in supplemental Sec.\,\ref{sec:combinatorics},
\begin{equation}
\Omega\!\left(N_{R}\right) = \binom{\ell-N_{R}}{N-N_{R}} \binom{\ell-N+N_{R}}{N_{R}}.
\label{eq:microstate_number}
\end{equation}
Notice that the two factors are invariant under flipping $L$ and $R$, 
mapping $N_{R}$ to $N-N_{R}$ and vice versa.
Intuitively, the first binomial factor represents the number of ways $N-N_{R}$ $L$-linkers
can be placed in an overlap with $\ell$ sites, 
when $N_{R}$ of those sites are excluded by $R$-linkers.
The second factor is simply the symmetric counterpart to the first one, and counts permutations of the $R$-linkers. 
The product of binomial coefficients is relatively small for $N_{R}=0$ or $N_{R}=N$, 
but reaches a peak around $N_{R}=N/2$,
where there are more possibilities due to permutations of left- and right-pulling linkers among each other.
With Eq.\,\ref{eq:potential_energy} and Eq.\,\ref{eq:microstate_number}, 
we have arrived at an exact solution for the free-energy Eq.\,\ref{eq:helmholtz_free_energy}.
\begin{figure}
	\includegraphics{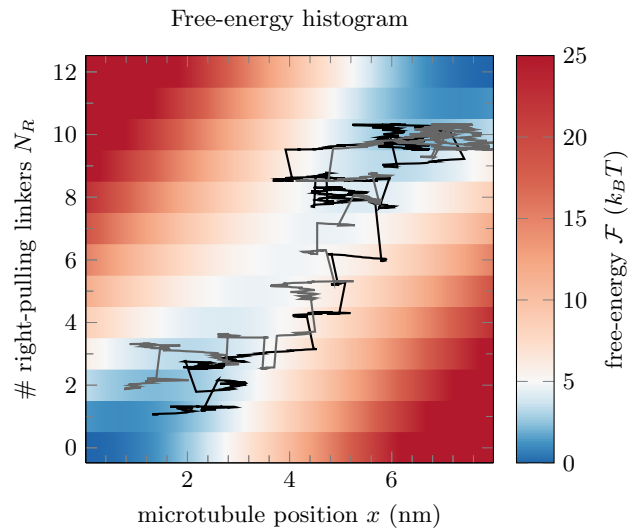}
	\caption{\label{fig:analytical_2d_histogram}
		Helmholtz free-energy as a function of the position $x$ 
		and the number of right-pulling crosslinkers $N_{R}$. 
		Free-energy minima exist at the bottom-left and top-right corners,
		which correspond to two neighboring aligned states of the microtubule.
		A higher saddle point exists in the center, which forms a pass connecting the two minima.
		A transition over the saddle point is observed as a jump of the microtubule,
		and two transition paths are shown for illustration.
		$N_{R}$ is a discrete parameter, 
		and an oscillating y-offset was added to the paths to visualize their course.
		In this example, we use $N=12$ and $\ell=40$.}
\end{figure}
We also checked this directly by comparing simulation results with the analytical free-energy equation
in supplemental Sec.\,\ref{sec:numerical_confirmation_free_energy}.

We plot the exact free-energy as a function of the order parameters in Fig.\,\ref{fig:analytical_2d_histogram},
which clearly shows that a free-energy barrier exists between two minima.
Hence, the jumping behavior can be understood as a barrier crossing phenomenon,
and the microtubule has to overcome the low probability region every time it makes an $\SI{8}{\nano\metre}$ move
over one tubulin dimer.

\section{\label{sec:arrhenius} Microtubule jump rate follows Arrhenius' law}
\begin{figure}
	\includegraphics{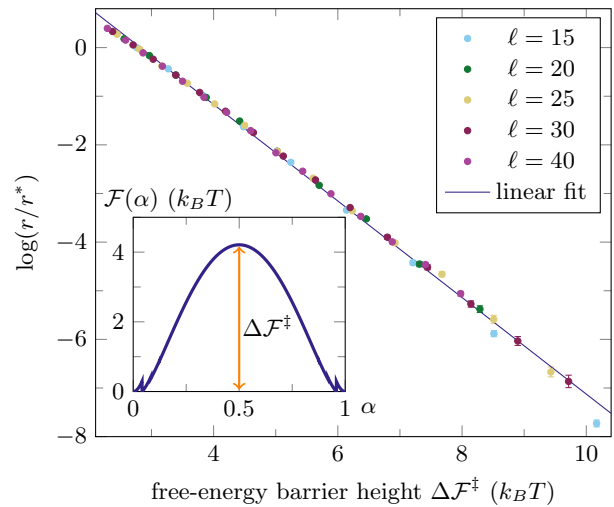}
	\caption{\label{fig:log_rate_vs_barrier_height_inset_alpha_barrier}
		Microtubule jump rate approximately follows Arrhenius' equation.
		We plot the natural logarithm of the simulated microtubule jump rate
		as a function of the height of the free-energy barrier separating two basins of attraction.
		The rate is normalized by an arbitrarily chosen rate $r^{*}$.
		We vary the barrier height by changing both $N$ and $\ell$,
		showing that all points fall on a single master curve independent of these parameters.
		A linear fit gives a slope of $\SI{-0.99}{\per k_{B}T}$,
		which shows that the rates are well described by Arrhenius' equation.
		(inset) The free-energy profile as a function of the reaction coordinate $\alpha$.
		The height of the barrier is the difference between the Helmholtz free-energies at $\alpha=0$ and $\alpha=1/2$.
		Discontinuities occur due to the discrete nature of $N_{R}$ in the definition of $\alpha$.
		}
\end{figure}

The height of the free-energy barrier directly influences the microtubule jump rate,
and we are interested in how this rate depends on 
the number of crosslinkers $N$ and the length of the mobile microtubule $\ell$.
To calculate the height of the barrier, we first need to find a proper reaction coordinate.
This will allow us to map the free-energy profile to a one dimensional one,
and to find the effective barrier that limits the microtubule dynamics.
As can be seen in Eq.\,\ref{eq:potential_energy}, the lowest free-energy path that connects 
the point $(x=0, N_{R}=0)$ with $(x=\delta, N_{R}=N)$ obeys $x/\delta = N_{R}/N$,
which corresponds to the diagonal of Fig.\,\ref{fig:analytical_2d_histogram}.
We introduce a reaction coordinate for the whole system that monotonically follows this path,
\begin{equation}
\alpha = \frac{1}{2} \left(\frac{x}{\delta} + \frac{N_{R}}{N}\right).
\label{eq:reaction_coordinate}
\end{equation}
This parameter increases from $0$ to $1$ along the bottom-left to top-right diagonal 
in Fig.\,\ref{fig:analytical_2d_histogram}.
As shown in supplemental Sec.\,\ref{sec:choice_reaction_coordinate}, Fig.\,\ref{fig:transition_histogram}, 
and Fig.\,\ref{fig:committer_histogram},
transition paths typically follow this diagonal, and the transition state ensemble is perpendicular to it.
Here, the transition state ensemble is defined as the collection of states 
that have the highest probability of being on a transition path~\cite{hummer_transition_2003}.
These states approximately coincide with the region constrained by $\alpha=1/2$.
Therefore, $\alpha$ is a proper reaction coordinate, 
and the free-energy profile can be calculated as a function of $\alpha$,
\begin{equation}
\mathrm{e}^{-\beta \mathcal{F}\!\left(\alpha\right)} = \sum_{N_{R}=0}^{N} \int_{0}^{\delta} \!\!
\mathrm{e}^{-\beta\mathcal{F}\left(x,N_{R}\right)}
\delta\!\left(\alpha\!\left(x,N_{R}\right)-\alpha\right) \diff{x}, % no space decrease in function seems better
\label{eq:free_energy_alpha}
\end{equation}
where $\delta\!\left(y\right)$ is the Dirac delta function 
and $\alpha\!\left(x,N_{R}\right)$ is the function given by Eq.\,\ref{eq:reaction_coordinate}.
As illustrated in the inset of Fig.\,\ref{fig:log_rate_vs_barrier_height_inset_alpha_barrier},
this free-energy profile sets the effective barrier height separating the two regions of attraction,
\begin{equation}
\Delta \mathcal{F}^{\ddagger} = \mathcal{F}\!\left(\alpha=1/2\right)-\mathcal{F}\!\left(\alpha=0\right).
\label{eq:barrier_height}
\end{equation}

We can use the free-energy barrier height to test if the reaction rate follows 
Arrhenius' equation~\cite{arrhenius_uber_1889},
\begin{equation}
r\!\left(x,N_{R}\right) = r_{0} \exp \!\left(-\beta \Delta \mathcal{F}^{\ddagger}\!\left(x,N_{R}\right)\right).
\label{eq:arrhenius}
\end{equation}
We confirm its validity by performing prolonged kinetic Monte Carlo simulations of the model,
varying $\ell$ between $15$ and $40$ sites, and varying $N$ between $6$ and, respectively, 
$14$ ($\ell=15$), $15$ ($\ell=20$), $18$ ($\ell=25$), and $20$ ($\ell=30$ and $\ell=40$) linkers.
We record the times $\tau$ between barrier crossings and estimate the hopping rate from the mean waiting time,
as illustrated in Fig.\,\ref{fig:position_vs_time_inset_transition_and_rate_vs_N}.
Then, we normalized the rates by dividing by some $r^{*}$,
for which we arbitrarily chose the empirically determined rate found at $\ell=15$ and $N=6$.
We plot the logarithm of these empirically obtained rates against the analytically calculated 
free-energy barrier height in Fig.\,\ref{fig:log_rate_vs_barrier_height_inset_alpha_barrier}.
This plot shows that all observed rates fall on a single master curve,
which is correctly described by Eq.\,\ref{eq:arrhenius} with a constant prefactor.
Therefore, we will use Eq.\,\ref{eq:arrhenius} and the analytical expression for the free-energy to predict 
how the microtubule jump rate depends on the parameters $N$ and $\ell$ in general. 

\section{\label{sec:prediction_jump_rate} Jump rate decreases superexponentially}
According to Eq.\,\ref{eq:einstein_relation}, the friction coefficient is determined by the microtubule jump rate,
and we showed in the previous section that 
this rate follows Arrhenius' Eq.\,\ref{eq:arrhenius} with a constant prefactor.
Therefore, to understand how the friction coefficient depends on 
the number of crosslinkers $N$ and the number of sites in the overlap $\ell$ 
we require the free-energy barrier height as a function of these parameters.
Our expression for the free-energy 
Eqs.\,\ref{eq:potential_energy},\ref{eq:microstate_number},\ref{eq:barrier_height} is exact, 
but it remains obscure how it is shaped by $N$ and $\ell$
due to the expression of the entropic term in terms of discrete binomial coefficients.
To reveal these parameter dependences, we make a continuous approximation 
to the expression for the free-energy barrier height.
First, we apply the Stirling approximation to the entropic term,
and approximate the binomial coefficients based on a well known procedure~\cite{milewski_derivation_2007}.
Further, we use a Gaussian integral to resolve the summation from Eq.\,\ref{eq:free_energy_alpha},
and we perform a Taylor expansion to keep only leading terms in $N$ and $N/\ell$.
Finally, we group the terms depending on the crosslinker density $N/\ell$ into a separate factor
and exponentiate this factor, since plots suggest an exponential behavior.
We stress that this procedure gives an analytical expression for the exponent, and requires no data fitting.
The full derivation of the free-energy approximation 
can be found in supplemental Sec.\,\ref{sec:approximation_free_energy}.
As a result, we find the following simplified expression for the barrier height,
\begin{equation}
\beta \Delta \mathcal{F}^{\ddagger} \approx A + B N \exp\!\left(\frac{1}{4B} \frac{N}{\ell}\right),
\label{eq:barrier_height_approximation_result}
\end{equation}
where
\begin{equation}
A = \frac{1}{2} \log \! \left(1+\frac{3 k \delta^{2}}{4 k_{B}T}\right), 
\quad B = \frac{k \delta^{2}}{8 k_{B}T} - \log\!\left(2\right).
\label{eq:approximation_parameters}
\end{equation}
These parameters are positive since $k$ is large relative to $k_{B}T/\delta^{2}$, 
making energetic contributions larger than entropic ones.
Now, Eq.\,\ref{eq:barrier_height_approximation_result} and Eq.\,\ref{eq:arrhenius} 
allow us to make predictions about how the friction coefficient changes 
with the number of crosslinkers $N$ or the microtubule overlap length $\ell$.
\begin{figure}
	\includegraphics{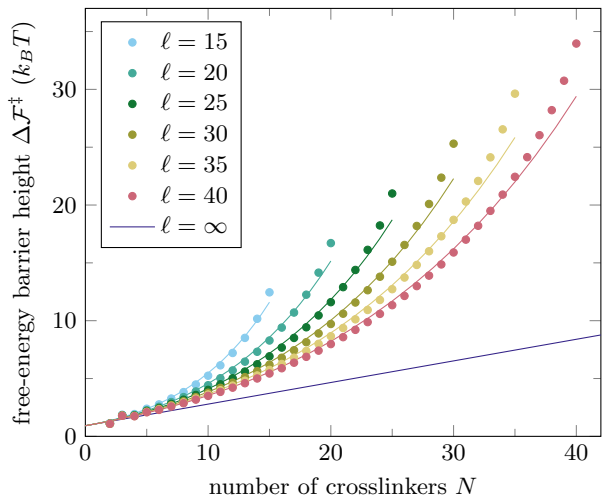}
	\caption{\label{fig:barrier_height_vs_n}
		The free-energy barrier height $\Delta\mathcal{F}^{\ddagger}$ increases exponentially with
		the number of crosslinkers $N$.
		The exact values (given by Eq.\,\ref{eq:barrier_height} and Eq.\,\ref{eq:rewrite_partition_sum_barrier}),
		plotted as points, are approximated well by the continuous exponential curves
		as given by Eq.\,\ref{eq:barrier_height_approximation_result}.
		Notice that the number of crosslinkers cannot exceed the number of sites on the microtubule, $N\leq \ell$.
		Furthermore, we plot the approximated barrier height for an infinitely long mobile microtubule, 
		which demonstrates that the barrier height increases linearly when the crosslinker density is negligible.
		Using Eq.\,\ref{eq:arrhenius}, we predict that the microtubule jump rate 
		decreases exponentially with $N$ for small densities,
		and decreases superexponentially with the density $N/\ell$.
	}
\end{figure}

Fig.\,\ref{fig:barrier_height_vs_n} shows how the barrier height depends on $N$ and $\ell$,
for both the exact and approximated results given by 
Eq.\,\ref{eq:barrier_height} and Eq.\,\ref{eq:barrier_height_approximation_result} 
(and supplemental Eq.\,\ref{eq:rewrite_partition_sum_barrier} and Eq.\,\ref{eq:free_energy_barrier_approximation}).
The exponential approximation of the barrier height is in surprisingly good agreement with the exact results,
as demonstrated in supplemental Fig.\,\ref{fig:barrier_height_vs_n_approximations}.
For very low crosslinker densities, $N/\ell\ll1$, the barrier height increases linearly with $N$.
There, a microtubule jump requires independent hops from all crosslinkers
and simultaneous microtubule movements in the same direction, 
roughly following the diagonal in Fig.\,\ref{fig:analytical_2d_histogram}.
These two actions can occur in any order,
but the net effect is that the system transitions through $N$ independent hops.
Each has some fixed average probability $q$ of occurring,
and this probability is relatively low because of the stiffness of the springs.
Therefore, the microtubule jump rate is proportional to $r \propto q^{N}$.
Hence, the rate decreases exponentially with $N$, 
and the barrier height found in Eq.\,\ref{eq:arrhenius} increases linearly.
Eq.\,\ref{eq:barrier_height_approximation_result} shows that $q=\exp\!\left(-B\right)$.
Then, for higher densities, exclusion effects begin to increase the friction,
since some crosslinker hops will be blocked.
This latter effect causes the barrier height to increase exponentially with $N/\ell$,
manifesting itself as a superexponential decrease of the microtubule jump rate.

Using the Einstein relation Eq.\,\ref{eq:einstein_relation},
we are able to predict how the friction coefficient depends on both $N$ and $\ell$.
We observe that friction increases superexponentially, 
\begin{equation}
\zeta \propto \exp\!\left(BN \exp\!\left(\frac{1}{4B} \frac{N}{\ell}\right)\right).
\label{eq:superexponential_friction}
\end{equation}
Consequently, the friction coefficient is hypersensitive to the number of crosslinkers in the overlap region
and to the size of the overlap.
Specifically, Eq.\,\ref{eq:superexponential_friction} shows that the friction increases rapidly 
when the overlap length drops below $N \delta/4B$.
For the parameter values listed in supplemental Tab.\,\ref{tab:parameter_values} 
and $N$ in the range $10 - 100$, our analysis predicts that 
this critical overlap length is around $\sim 0.1-1\si{\micro\metre}$.

\section{\label{sec:discussion} Discussion}
Here, we used a simple model to show that friction between microtubules caused by diffusible crosslinkers 
increases non-linearly with the number of crosslinkers in the overlap.
This model was previously successful in explaining 
the entropic forces arising in microtubule overlaps~\cite{lansky_diffusible_2015},
and the combination of the model and experiments suggested an exponential increase in the friction coefficient.
Here, we exactly solved the free-energy landscape of the model 
and found that microtubule movement is limited by barrier crossings.
We identify the effective spring constant of crosslinkers, the number of crosslinkers, 
and the microtubule overlap length as key parameters that set the height of the barrier.
The latter two parameters are experimentally accessible,
and can be used to test our model predictions.
We observe that for very low crosslinker densities, 
the friction coefficient for microtubule sliding increases approximately exponentially,
whereas friction increases superexponentially at finite crosslinker densities.

The non-linearity of the friction coefficient crucially depends on 
the discrete nature of the microtubule binding sites for crosslinkers.
Friction would scale linearly with the number of crosslinkers 
if crosslinkers moved over the microtubules in a continuous motion,
with no preferential binding sites.
Hence, the dependence of the friction coefficient on the crosslinkers is an experimentally accessible characteristic
that can distinguish between two models of crosslinker binding.
New experiments which directly measure either the diffusion constant of microtubules or their friction coefficient
could test whether the friction coefficient increases superexponentially with the crosslinker density.
The predicted scaling, Eq.\,\ref{eq:superexponential_friction}, only contains a single fit parameter, 
and we have estimated this constant from previous experiments already
as shown in supplemental Sec.\,\ref{sec:estimate_spring_constant}.
We predict that changing the crosslinker density in the overlap from $10\%$ to $40\%$ 
would increase the friction coefficient by roughly two orders of magnitude.
Hence, the effect should be clearly observable in experiments.

Besides providing specific predictions on microtubule friction, 
the model grants opportunities for studying barrier crossings in general and to test reaction-rate theories.
The model dynamics is relatively simple, yet it still shows emerging barrier crossings.
Since we also found an exact solution of the free-energy profile for this transition,
it is interesting to study theories that predict the rate of barrier crossings from this free-energy landscape.
Specifically, it would be of interest in future work to investigate whether
the prefactor of the microtubule jump rate can be described by Kramers' theory~\cite{kramers_brownian_1940}.
Also, it would be interesting to study the influence that parameters such as 
the bare crosslinker hopping rate $h_{0}$ or the microtubule diffusion constant $D_{M}$
have on the transition state and optimal reaction coordinate.
It is possible that transition paths would no longer follow the optimal free-energy path
when the timescales of crosslinker and microtubule dynamics differ significantly~\cite{ten_wolde_drying-induced_2002}.

The superexponential dependence of the friction on the crosslinker density has implications in biology, 
mainly in the control of the overlap region.
During cell division, microtubules in the mitotic spindle overlap 
and are pushed apart by plus-end directed motor proteins.
Since the friction coefficient is hypersensitive not only to the number 
but also to the density of proteins crosslinking two microtubules,
we predict that a shrinking overlap region will undergo a sudden increase in the friction coefficient
on timescales where crosslinker unbinding can be neglected.
This will effectively stall the microtubule,
and impose a precise overlap length in the midzone where the opposing spindle microtubules meet.
We estimate the size of the stationary overlap region to be on the order of $0.1-1\si{\micro\metre}$, 
which appears reasonable given the size of the spindle midzone~\cite{jiang_prc1:_1998,mollinari_prc1_2002}.
This length scale can be fine-tuned by controlling the number of crosslinkers contained in the overlap,
for example by reducing the binding affinity of PRC1 to microtubules 
through phosphorylation of a microtubule binding domain on PRC1~\cite{kellogg_near-atomic_2016}.

\begin{acknowledgments}
	The authors thank Z. Lansky, M. Braun, and S. Diez for the fruitful collaboration,
	and B.M. Mulder for assessing the manuscript.
	This work was supported by European Research Council (ERC) Synergy Grant 609822,
	is part of the research programme of the Netherlands Organisation for 
	Scientific Research (NWO), and performed at the research institute AMOLF.
\end{acknowledgments}


% MERGE WITH SUPPLEMENTAL MATERIAL

\widetext
\pagebreak
\begin{center}
	\textbf{\large Supplemental Material: \\ 
		Diffusible crosslinkers cause non-linear friction forces}
\end{center}

\setcounter{equation}{0}
\setcounter{figure}{0}
\setcounter{table}{0}
\setcounter{page}{1}
\setcounter{section}{0}
\makeatletter
% Put an "S" before all equation numbers
\renewcommand{\theequation}{S.\arabic{equation}}
\renewcommand{\thesection}{S.\Roman{section}}
\renewcommand{\thefigure}{S.\arabic{figure}}
\renewcommand{\thetable}{S.\arabic{table}}
\renewcommand{\bibnumfmt}[1]{[S#1]}
\renewcommand{\citenumfont}[1]{S#1]}

%======================================================================
%
\begin{table}[b]
	\begin{ruledtabular}
		\begin{tabular}{lll}
			Crosslinker hopping rate $h_{0}$ & \SI{1562.5}{\per\second}\\
			Spring constant $k$ & \SI{1.1e5}{k_{B}T\per\micro\meter\squared}\\
			Diffusion constant bare microtubule $D_{m}$ & \SI{0.01}{\micro\meter\squared\per\second}\\
			Lattice spacing binding sites $\delta$ & \SI{0.008}{\micro\metre}
		\end{tabular}
	\end{ruledtabular}
	\caption
	{\label{tab:parameter_values} 
		Model parameters and values.
		The base hopping rate $h_{0}$ was fit to the observed diffusion constant $D_{s}$ of Ase1 on single microtubules,
		and subsequently $k$ was fit to the observed diffusion constant $D_{d}$ between two microtubules
		(see Sec.\,\ref{sec:estimate_spring_constant}).
		$D_{m}$ was estimated from previously reported experiments (see Sec.\,\ref{sec:simulation_dynamics}).
		$\delta$ approximately equals the tubulin dimer length.
	}
\end{table}
\section{\label{sec:estimate_spring_constant} Estimate of spring constant revisited}
The model parameters were previously estimated from experimental data~\cite{sup_lansky_diffusible_2015}.
Specifically, the spring constant was estimated by measuring the diffusion constant of linkers on a single microtubule
$D_{s}$ (s for singly bound),
and the diffusion constant of crosslinkers in an overlap region $D_{d}$ (d for doubly bound).
There, measurement of the mean square displacement as a function of time gave values of
$D_{s}=\SI{0.085(7)}{\micro\metre\squared\per\second}$ and $D_{d}=\SI{0.011(3)}{\micro\metre\squared\per\second}$.
Then, the spring constant was fitted by varying it in simulations until the simulated diffusion constant 
roughly matched the measured one, giving a value of $k=\SI{1.1e5}{k_{B}T\per\micro\meter\squared}$.
Here, we present an analytical expression for the spring constant in terms of the measured diffusion constants.

First, consider the diffusion of a single linker on one microtubule.
We assume that the hopping rate of one head equals $h_{0}$.
Then, when the lattice spacing $\delta=\SI{8}{\nano\metre}$ is taken into account, 
we get the diffusion constant $D_{s}=h_{0} \delta^{2}$.
For consistency with previous work, we use $h_{0}=\SI{1562.5}{\per\second}$~\cite{sup_lansky_diffusible_2015}.

\afterpage{\begin{figure}
	\includegraphics{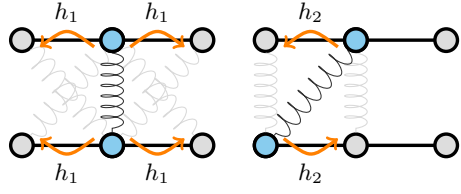}
	\caption{\label{fig:linker_diffusion_constant_overlap}
		Possible transitions for a diffusing crosslinker in an overlap region.
		For the estimation of the diffusion constant, we assume that the two microtubules stay fixed relative to each other,
		and that they are perfectly aligned.
		Then, crosslinkers can be either in a relaxed straight (left) or stretched diagonal (right) state.
		When we only focus on transitions towards one direction, e.g. the left, 
		then there are two possible transitions from the straight position,
		whereas there is only one possible transition from the diagonal state. 
	}
\end{figure}}
The diffusion of a doubly bound crosslinker in an overlap region is more complicated, 
as several different transitions are possible.
We assume that the microtubule does not move (much) during this diffusion, 
since the time scale of microtubule movement is much longer than that of individual crosslinker hopping.
Then, the crosslinker can be either in a straight or in a diagonal state, 
as shown in Fig.\,\ref{fig:linker_diffusion_constant_overlap}.
Ignoring the finite overlap length, we label the straight states with even numbers, and the diagonals with odd numbers.
For example, a transition from state $0$ to state $1$ represents a hop of a straight linker to a diagonal on the right.
Both possible diagonal configurations, 
which are indicated in the left illustration of Fig.\,\ref{fig:linker_diffusion_constant_overlap}, 
are grouped into one state. 
We can call this number $Z\!\left(t\right) \in \mathbb{Z}$, 
and it relates to the physical position of the center of the crosslinkers $x$ through 
\begin{equation}
x\!\left(t\right)= \frac{\delta}{2} Z\!\left(t\right).
\label{eq:relation_x_Z}
\end{equation}
$Z\!\left(t\right)$ defines a Markov chain on the integers (a two-sided birth-death process),
with probabilities $p_{n}\!\left(t\right)$ to be in state $n$ at time $t$.
The time derivatives of these probabilities are set by the transition rate to leave a straight conformation, $h_{1}$,
and the rate to leave a diagonal conformation, $h_{2}$.
These time derivatives are different for the straight state probabilities ($n=2m$) 
and diagonal state probabilities ($n=2m+1$),
\begin{align}
\partial_{t} p_{2m} &= -2 h_{1} p_{2m} + h_{2} p_{2m-1} + h_{2} p_{2m+1} \nonumber \\*
\partial_{t} p_{2m+1} &= -2 h_{2} p_{2m+1} + h_{1} p_{2m} + h_{1} p_{2m+2}.
\label{eq:time_derivatives_probabilities_j}
\end{align}
The hopping rates equal
\begin{align}
h_{1} &= 2 h_{0} \exp\!\left(-\frac{1}{4} k \delta^{2}\right) \nonumber \\*
h_{2} &= h_{0} \exp\!\left(\frac{1}{4} k \delta^{2}\right),
\label{eq:hopping_rates}
\end{align}
where $h_{0}$ is the rate prefactor.
We can calculate the long time diffusion constant of the hopping crosslinkers from Eq.\,\ref{eq:time_derivatives_probabilities_j}.
To that end, we look at the mean square displacement and take its time derivative,
\begin{align}
\partial_{t} \langle Z\!\left(t\right)^{2} \rangle &= \partial_{t} \sum_{n=-\infty}^{\infty} n^{2} p_{n}\!\left(t\right)
= \sum_{m=-\infty}^{\infty} \bigg\{ (2m)^{2} \partial_{t} p_{2m}\!\left(t\right) 
+ (2m+1)^{2} \partial_{t} p_{2m+1}\!\left(t\right) \bigg\} \nonumber \\*
&= \sum_{m=-\infty}^{\infty} \bigg\{ (2m)^{2} \left[-2 h_{1} p_{2m} + h_{2} p_{2m-1} + h_{2} p_{2m+1}\right] \nonumber \\*
&\qquad +(2m+1)^{2} \left[-2 h_{2} p_{2m+1} + h_{1} p_{2m} + h_{1} p_{2m+2}\right] \bigg\} \nonumber \\*
&= \sum_{m=-\infty}^{\infty} \bigg\{ \left[-2(2m)^{2} + (2m+1)^{2} + (2m-1)^{2} \right] h_{1} p_{2m} \nonumber \\*
&\qquad + \left[-2(2m+1)^{2} + (2m+2)^{2} + (2m)^{2} \right] h_{2} p_{2m+1} \bigg\} \nonumber \\*
&= 2 h_{1} \!\! \sum_{m=-\infty}^{\infty} p_{2m} + 2 h_{2} \!\! \sum_{m=-\infty}^{\infty} p_{2m+1}.
\label{eq:time_derivative_mean_square_displacement}
\end{align}
Then, we use that in steady state, the probability to be in an even or in an odd state is
\begin{align}
\sum_{m=-\infty}^{\infty} p_{2m} &= \frac{h_{2}}{h_{1}+h_{2}}, \nonumber \\*
\sum_{m=-\infty}^{\infty} p_{2m+1} &= \frac{h_{1}}{h_{1}+h_{2}}.
\label{eq:steady_state_probabilities}
\end{align}
Then we combine Eq.\,\ref{eq:time_derivative_mean_square_displacement} and Eq.\,\ref{eq:steady_state_probabilities},
together with its relation with the mean square displacement of the physical position through Eq.\,\ref{eq:relation_x_Z},
\begin{equation}
\partial_{t} \langle x\!\left(t\right)^{2} \rangle = \delta^{2} \frac{h_{1} h_{2}}{h_{1}+h_{2}} = 2 D_{d}.
\label{eq:mean_square_displacement_diffusing_linker}
\end{equation}
The final equality follows from the standard relation between the mean square displacement in one dimension 
and the diffusion constant.
Then we can insert the definitions of the rates, Eq.\,\ref{eq:hopping_rates}, and solve the equation for the spring constant $k$.
This gives us
\begin{equation}
k = \frac{2}{\delta^{2}} \left[\log\!\left(2\right) +2 \arccosh\!\left(\frac{D_{s}}{2\sqrt{2} D_{d}}\right) \right].
\label{eq:expression_k}
\end{equation}
Here, we used the simple relation $D_{s}=\delta^{2} h_{0}$ to rewrite the result in terms of the diffusion constant.
Evaluating this expression for the experimentally measured quantities gives $k=\SI{1.26e5}{k_{B}T\per\micro\metre\squared}$, 
which is indeed very close to the computationally estimated $k=\SI{1.1e5}{k_{B}T\per\micro\metre\squared}$.
For consistency with the previously published work~\cite{sup_lansky_diffusible_2015}, 
we keep using the value reported there in the rest of this work.
A summery of the model parameters is given in Tab.\,\ref{tab:parameter_values}.

%======================================================================
\section{\label{sec:simulation_dynamics} Simulation dynamics}
Crosslinkers exert pulling forces on the microtubule, and the net force $F$ cause it to move.
We only allow the microtubule to move in one dimension, 
and consider the drag between the mobile microtubule and the fluid to be in the overdamped regime.
Hence, we can model the time evolution of the microtubule position $x$ using Brownian dynamics, 
\begin{equation}
\frac{\diff x}{\diff t} = \frac{F}{\gamma_{m}} + \eta.
\label{eq:brownian_dynamics}
\end{equation}
Here, the $\gamma_{m}$ is the drag coefficient of the mobile microtubule when it is not linked by crosslinkers,
and $\eta$ is the thermal noise.
The noise amplitude is set by the bare microtubule diffusion constant $D_{m}$,
and is related to $\gamma_{m}$ through the Einstein relation~\cite{sup_einstein_uber_1905}, 
\begin{equation}
\gamma_{m} = \frac{k_{B}T}{D_{m}}.
\label{eq:einstein_relation_bare_microtubule}
\end{equation}
Previously, the diffusion constant was estimated~\cite{sup_lansky_diffusible_2015} 
from a formula for the parallel translational drag coefficient~\cite{sup_hunt_force_1994},
which gives us $D_{m} = \SI{0.01}{\micro\meter\squared\per\second}$ independent of microtubule length.
In simulations, we assume $F$ to be constant during finite time steps of size $\Delta t$, 
and update the microtubule position with the deterministic term and a Gaussian noise term,
\begin{equation}
\Delta x = \frac{D_{m}}{k_{B}T} F \Delta t + \sqrt{2 D_{m} \Delta t} \ \mathcal{N}.
\label{eq:simulation_position_update}
\end{equation}
Here, $\mathcal{N}$ denotes a Gaussian random variable with zero mean and unit standard deviation.

The net force on the microtubule $F$ depends on the full system state, 
so both on the position $x$ and on the number of right-pulling crosslinkers $N_{R}$.
Furthermore, when some crosslinkers get close to their maximum stretch of $\delta$,
not all values of $\Delta x$ are allowed.
The maximum stretch of the crosslinkers is taken into account by placing reflective boundary conditions on $\eta$.
We implement the algorithm by first updating the position using the deterministic term,
since deterministic change never causes $\Delta x$ to cross a boundary imposed by the crosslinkers.
Then, we calculate the boundary $b$ for the stochastic term $\mathcal{N}$, 
and if $\mathcal{N}$ passes $b$, we reflect the term through
\begin{equation}
\mathcal{N} \rightarrow 2b - \mathcal{N}.
\label{eq:reflection_noise}
\end{equation}

For the crosslinker dynamics, we use a kinetic Monte Carlo algorithm~\cite{sup_prados_dynamical_1997}
to simulate the hops.
Each possible hop $i$ has some time-varying rate $h_{i}\!\left(x,t\right)$ of occurring, 
as shown in Eq.\,\ref{eq:rate_h_form} of the main text.
Then, we calculate the full rate 
\begin{equation}
H\!\left(t\right) = \sum_{i} h_{i}\!\left(x,t\right).
\end{equation}
To decide the next moment at which a reaction will take place,
we draw a uniform random number $\xi$ between $0$ and $1$, which represents a survival probability $S\!\left(t\right)$,
for which
\begin{equation}
S\!\left(t\right) = \exp\!\left(-\int_{0}^{t}\diff{t'} H\!\left(t'\right) \right).
\label{eq:survival_probability}
\end{equation}
We choose the time steps small enough to keep the rate $h\!\left(x,t\right)$ nearly constant between steps,
such that we can approximate the time integral as a sum over time steps,
\begin{equation}
\int_{0}^{t}\diff{t'} H\!\left(t'\right) \approx \sum_{n=1}^{t/\Delta t} H\!\left(n \Delta t\right) \Delta t.
\label{eq:action_as_sum}
\end{equation}
In the simulations, we update the integral after each time step,
and a hop is performed when the integrated total rate reaches a threshold value given by $\xi$,
\begin{equation}
\sum_{n=1}^{t/\Delta t} H\!\left(n \Delta t\right) > -\frac{\log\!\left(\xi\right)}{\Delta t}
\label{eq:threshold}
\end{equation}
After the hop, a new value of $\xi$ is drawn and the integral is reset to $0$.

%======================================================================

\section{\label{sec:microtubule_rate} Microtubule jumps are Markovian}
\begin{figure}
	\includegraphics{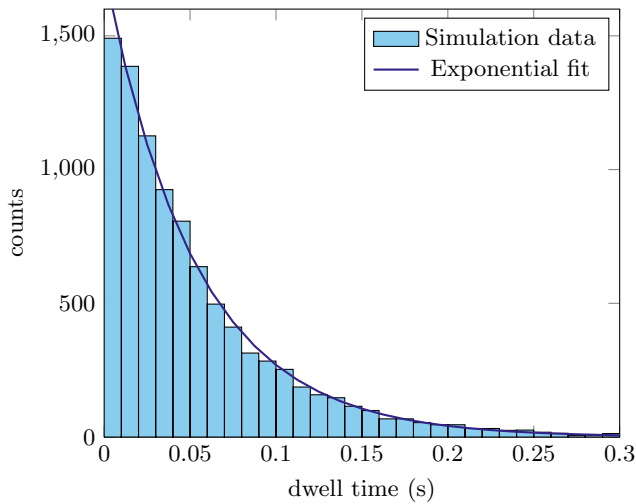}
	\caption{\label{fig:dwell_time_hist}
		Histogram of the dwell times $\tau$ at certain lattice positions, as observed for a system with $N=12$ and $\ell=40$.
		The red line shows an exponential distribution with rate $r=1/\overbar{\tau}$.}
\end{figure}
\begin{figure}
	\includegraphics{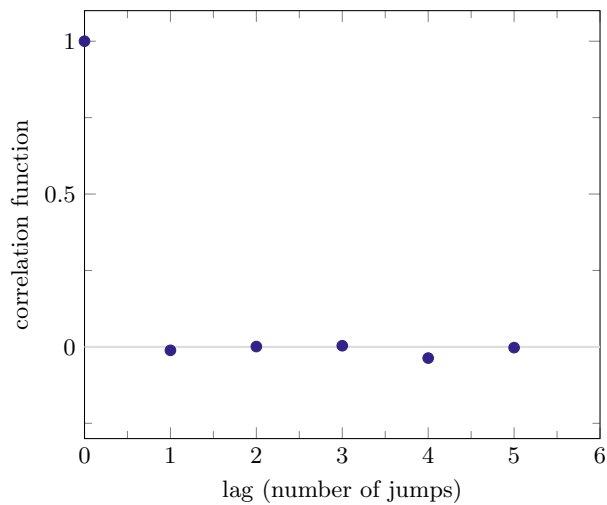}
	\caption{\label{fig:jump_correlation_vs_lag}
		Autocorrelation function of the jump direction. Time is represented by the number of stochastic jumps in between.
		We observe no correlation in the direction of subsequent jumps.}
\end{figure}
The mobile microtubule moves in a discrete fashion on macroscopic time scales, 
as indicated in Fig.\,\ref{fig:position_vs_time_inset_transition_and_rate_vs_N} in the main text.
The filament makes jumps of size $\delta$, which equals one tubulin dimer length, 
to the left or right at random points in time.
We are interested in describing these jumps as a memoryless process,
with a fixed rate of jumping in a random direction.
To convince ourselves that this process is indeed Markovian, 
we plot an example histogram of observed dwell times in Fig.\,\ref{fig:dwell_time_hist}. 
The histogram shows an exponential distribution, as is expected for a continuous time Markov chain.
We estimate the rate of jumping using the maximum likelihood estimator
\begin{equation}
r = \frac{1}{\overbar{\tau}} = \frac{n}{\sum_{i=1}^{n} \tau_{i}},
\label{eq:maximum_likelihood_estimator_rate}
\end{equation}
where $\tau_{i}$ are the $n$ different samples of the waiting time between two microtubule jumps.
Then, to compare the estimated exponential distribution to the histogram of dwell times, 
we need to calculate the expected number of counts in each bin. 
We call the bin width $\delta \tau$, 
which means that bin $i$ captures values of $\tau$ between ${(i-1)\, \delta \tau}$ and ${i \, \delta \tau}$.
Then, the number of expected counts in bin $i$ is
\begin{equation}
n_{i} = n \int_{(i-1)\, \delta \tau}^{i \, \delta \tau} r \mathrm{e}^{-rt} \diff{t} 
= 2 \sinh \!\left(\frac{r \delta \tau}{2}\right) \mathrm{e}^{-\left(i-\frac{1}{2}\right)r \delta \tau}
= 2 \sinh \!\left(\frac{r \delta \tau}{2}\right) \mathrm{e}^{-r \sigma}.
\label{eq:expected_bin_counts}
\end{equation}
We call ${\sigma=\left(i-\frac{1}{2}\right)\, \delta \tau}$ the continuous time axis of the histogram.
This choice ensures that the exponential fit coincides with the centers of bins.
As can be seen in Fig.\,\ref{fig:dwell_time_hist}, the histogram deviates little from the exponential fit.
For smaller barrier heights, e.g. for $\ell=40$ and $N=6$ with a barrier height of just ${\Delta\mathcal{F}^{\ddagger}=2.3 k_{B}T}$,
there starts to be a noticeable deviation for small times.
Specifically, for low barriers it becomes apparent that it takes time to cross the barrier and equilibrate in the valley,
which makes it less likely to quickly jump twice in a row.
For high barriers, the typical timescale between jumps is so long that the time it takes to cross the barrier is negligible.
Hence, at least for moderately high barriers, the jump times are memoryless.
Then, as shown in Fig.\,\ref{fig:jump_correlation_vs_lag}, also jump directions are memoryless.
There are no correlations between subsequent jump directions.
Therefore, we can treat the dynamics of the microtubule as an emergent Markov process with jump rate $r$ as its only parameter.

%======================================================================

\section{\label{sec:combinatorics} Configuration combinatorics for entropy calculation}
\begin{figure}
	\includegraphics{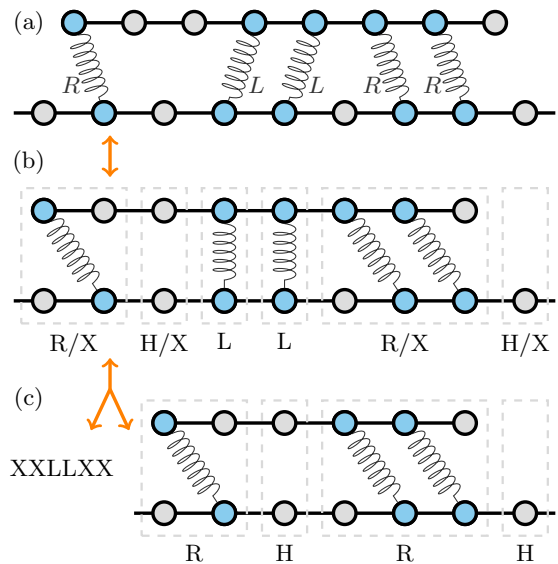}
	\caption{\label{fig:configuration_blocks}
		Map to calculate the number of configurations given the number of right-pulling crosslinkers $N_{R}$.
		Here, we show an example configuration with $\ell=8$, $N=5$, and $N_{R}=3$.
		(a) The number of ways to place $N_{R}$ $R$ linkers and $N-N_{R}$ $L$ linkers in the overlap 
		does not depend on the mobile microtubule position $x$.
		Hence, for the purpose of counting configurations, we can ignore the exact alignment of the microtubules
		and focus on the combinatorics of linker placement.
		(b) In the picture where $L$ linkers are straight, they always exclude one site on the top microtubule.
		However, groups of adjacent $R$ linkers always exclude one more site.
		It is helpful to draw boxes around each unit, which can be an $L$ linker ($L$-block),
		a hole ($H$-block), or a set of adjacent $R$ linkers ($R$-block).
		The last block cannot be an $L$-block.
		The number of $R$- and $H$-blocks together always equals $\ell+1-N+N_{R}$, 
		and we alternatively label those $X$-blocks, to contrast with $L$-blocks.
		(c) There are always $N-N_{R}$ $L$-blocks, 
		and removing all $L$-blocks leaves a system with the same number of $X$-blocks, since none of those merge.
		This means that each configuration can be split into an arrangement of $L$- and $X$-blocks, 
		and into a specific organization of the crosslinkers within the microtubule with all $L$-blocks removed.
		The arrows represent the bijective mapping between the set of linker configurations in $(a)$ and $(c)$.
	}
\end{figure}
We describe a transition of the system through the change of the relative microtubule position and 
the number of right-pulling crosslinkers.
These quantities act as our order parameters, 
where a transition means that the microtubule position changes by one lattice spacing $\delta$ and 
the number of right-pulling linkers changes by the total number of crosslinkers $N$.
To find a description for the transition rate, we require the free energy profile as a function of these order parameters.

In our model, the potential energy only depends on $x$ and $N_{R}$, 
and is not influenced by other details of the system configuration.
This lets us describe the entropy as the logarithm of the number of microstates $\Omega\!\left(x,N_{R}\right)$
at a given $x$ and $N_{R}$.
Furthermore, the number of microstates is independent of the position $x$, 
since the amount of stretch does not change the classification into left- and right-pulling linkers,
as shown in Fig.\,\ref{fig:configuration_blocks}.
Hence,
\begin{equation}
S\!\left(x,N_{R}\right) = k_{B} \log\!\left(\Omega\!\left(N_{R}\right)\right).
\label{eq:entropy_log_omega}
\end{equation}

To calculate $\Omega\!\left(N_{R}\right)$ we use the mapping depicted in Fig.\,\ref{fig:configuration_blocks}.
There, we decompose the microstate into a set of blocks that group together different sets of 
microtubule sites and crosslinkers.
Each left-pulling linker is placed in its own box spanning a single site called an $L$-block,
whereas all neighboring right-pulling linkers are grouped together into a single $R$-block.
The latter is done to take into account the variable number of sites excluded by right-pulling linkers
when they are alone or in contact with other diagonally placed linkers.
Finally, the remaining sites are called holes and grouped into $H$-blocks.
\begin{table}
	\begin{ruledtabular}
		\begin{tabular}{lll}
			Block type & Number of blocks & Number of excluded sites\\
			$L$ & $N-N_{R}$ & $N-N_{R}$\\
			$R$ & $m$ & $N_{R}+m$\\
			$H$ & $\ell+1-N-m$ & $\ell+1-N-m$\\
			$X=R \vee H$ & $\ell+1-N$ & $\ell+1-N+N_{R}$\\
			all & $\ell+1-N_{R}$ & $\ell+1$
		\end{tabular}
	\end{ruledtabular}
	\caption
	{\label{tab:number of blocks} 
		The number of blocks and number of sites excluded for each type of block.
		The number of $R$ blocks variates among configurations, and is called $m$ here.
		The $X$-blocks are a name for the ensemble of $R$- and $H$-blocks, 
		which are grouped together since their number is independent of the only variable $m$.
	}
\end{table}
In Tab.\,\ref{tab:number of blocks}, we calculate the number of blocks of each type 
by counting the number of sites that are occupied by each type.
The numbers of $R$- and $H$-blocks are variable, but it turns out that their sum is constant.
Therefore, we group these two types under a new block name, called $X$-blocks.
Each configuration of the microtubule has a unique representation as a permutation of 
$\ell+1-N$ $X$-blocks and $N-N_{R}$ $L$-blocks,
with $\ell+1-N_{R}$ blocks in total.
Additionally, the last of these blocks always needs to be an $X$-block, 
since $L$-blocks cannot occupy the final site where there is no site available on the mobile microtubule.
Hence, the number of ways to permute the $L$- and $X$ blocks among each other is
\begin{equation}
\Omega_{1}\!\left(N_{R}\right) = \frac{\left(\ell+1-N_{R}-1\right)!}{\left(\ell+1-N-1\right)!\left(N-N_{R}\right)!}
= \binom{\ell-N_{R}}{N-N_{R}},
\label{eq:permute_L_blocks}
\end{equation}
where the $-1$ terms are due to the final block always being an $X$.

$\Omega_{1}\!\left(N_{R}\right)$ captures all permutations of the left side of Fig.\,\ref{fig:configuration_blocks}(c),
which leaves us with calculating the number of permutations of the remaining blocks.
Since removing an $L$ block never merges two neighboring blocks together,
we can remove all straight crosslinkers without changing the order of $X$ blocks.
Finally, we see that we are left with $\ell-N+N_{R}$ sites on the mobile microtubule
with $N_{R}$ linkers bound to it.
Without considering the blocks, we know the number of ways these linkers can be placed,
\begin{equation}
\Omega_{2}\!\left(N_{R}\right) = \binom{\ell-N+N_{R}}{N_{R}}.
\label{eq:permute_R_linkers}
\end{equation}
Each of these configurations actually constitutes a unique permutation of $R$- and $H$-blocks. 
This permutation is then substituted into the the $X$ positions in the string of $L$- and $X$-blocks,
such that we are left with a unique configuration of all linkers.
Hence, the total number of configurations is simply the product
\begin{equation}
\Omega\!\left(N_{R}\right) = \Omega_{1}\!\left(N_{R}\right) \Omega_{2}\!\left(N_{R}\right) 
= \binom{\ell-N_{R}}{N-N_{R}}\binom{\ell-N+N_{R}}{N_{R}}.
\label{eq:number_of_configurations}
\end{equation}
This concludes our calculation of the entropy term Eq.\,\ref{eq:entropy_log_omega}.

%======================================================================

\section{\label{sec:numerical_confirmation_free_energy} Numerical confirmation of free energy equation}
\begin{figure}
	\includegraphics{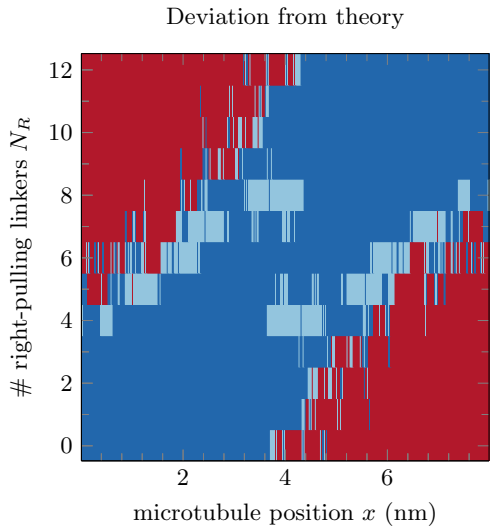}
	\caption{\label{fig:deviation_from_theory_histogram}
		Comparison of the exact free energy and a numerical estimate. 
		Dark blue colors represent a deviation from the theoretical value $d<2.5\%$,
		light blue represent $2.5\% \leq d < 5\%$, and red represents $d>5\%$.
		The numerical results confirm the validity of the equations in all regions that can be efficiently sampled.
		Here, $N=12$ and $\ell=40$, and the $x$ axis is divided into $400$ bins.
	}
\end{figure}

We test the analytical expression for the free energy profile by calculating it directly from numerical simulations.
We obtained a two dimensional histogram of $(x,N_{R})$ positions
by running simulations at $N=12$ and $\ell=40$ for $\num{5e10}$ time steps 
representing $\Delta t= \SI{1e-8}{\second}$ each. 
The barrier was crossed $9332$ times during this run, which shows we have sampled the peak region to some extend.
Then, we estimated the free energy by calculating $-\log(p)$, 
where $p$ is the probability to be found in a particular bin, 
and choosing a constant offset such that the free energy vanishes at $x=0$ and $N_{R}=0$.
The result is plotted in Fig.\,\ref{fig:deviation_from_theory_histogram}.
In regions where the free energy is relatively low, including the saddle point in the barrier,
the simulations confirm the exact free energy profile.
We did not sample regions with higher free energy due to the finite simulation time, 
thus showing a deviation from the theoretical result there.
Still, these simulations confirm the agreement between simulation and theory.
%

%======================================================================

\section{\label{sec:choice_reaction_coordinate} Determination of reaction coordinate}
As shown in the main text, two order parameters describe the microtubule jumps. 
First, the microtubule position $x$ modulo $\delta$ transitions between $0$ and $\delta$,
and second, the number of right-pulling crosslinkers $N_{R}$ ranges from $0$ to $N$.
As shown in Eq.\,\ref{eq:potential_energy} in the main text,
the lowest free energy path connecting two neighboring basins of attraction is given by the diagonal $x/\delta=N_{R}/N$.
\begin{figure}
	\includegraphics{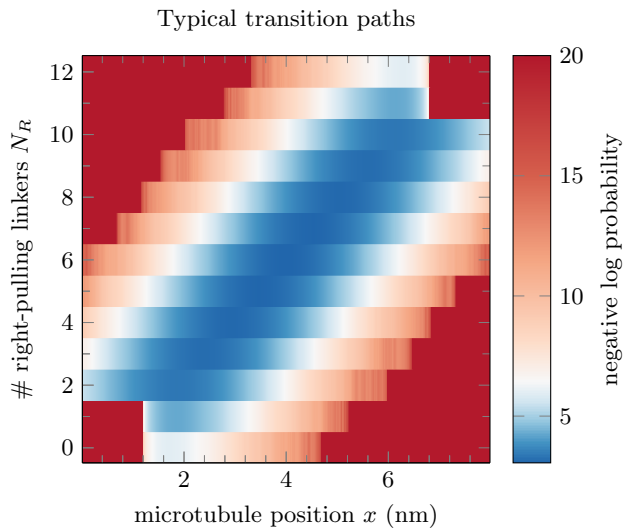}
	\caption{\label{fig:transition_histogram}
		Histogram of transition paths. 
		Two square corners of $\SI{1.2}{\nano\metre}$ wide and $2$ linkers high were set as the basins of attraction.
		Then we estimated the probability to be at a certain coordinate, 
		given that the system is on a transition path, $\mathcal{P}\!\left(x,N_{R}\mid \mathrm{transit} \right)$.
		This is simply a normalized histogram of all transition paths.
		Finally, we plot the negative natural logarithm of this probability.
		Most paths follow the diagonal, and it is unlikely to deviate far into the top-left or bottom-right corners,
		making it impossible to sample these latter regions.
		It is impossible to be part of a transition path in the basins of attraction,
		as shown by the squares in the bottom-left and top-right.
	}
\end{figure}
\begin{figure}
	\includegraphics{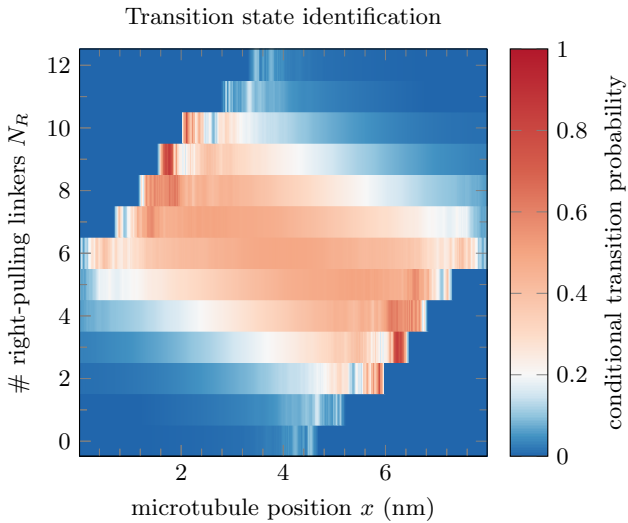}
	\caption{\label{fig:committer_histogram}
		Histogram of transition path probability, $\mathcal{P}\!\left(\mathrm{transit} \mid x,N_{R}\right)$. 
		The region of largest values, where $\mathcal{P}=1/2$,
		defines the transition state~\cite{sup_hummer_transition_2003}.
		Value fluctuations at the edges of the sampled region are due to undersampling,
		whereas the top-left and bottom-right corners are not sampled at all.
		The transition state is roughly perpendicular to the reaction coordinate, 
		and is approximately captured by $\alpha=1/2$.
	}
\end{figure}
The reaction coordinate $\alpha$ is defined such that a single value 
groups states perpendicular to the optimal diagonal,
\begin{equation}
\alpha = \frac{1}{2} \left(\frac{x}{\delta} + \frac{N_{R}}{N}\right).
\label{eq:reaction_coordinate_si}
\end{equation}
To study whether this gives the proper reaction coordinate,
we simulated a system with $N=12$ and $\ell=40$ 
and recorded the phase space positions $\left(x,N_{R}\right)$ at $\mathcal{N}_{max}$ points in time.
In this case, we used $\mathcal{N}_{max}=\num{5e10}$ time steps of $\Delta t = \SI{1e-8}{\second}$ each.
Then, we created a histogram with bin dimensions $\left(\SI{2.5e-3}{\delta},1\right)$ 
that collects those points that were part of transition paths. 
For this, the basins of attraction are defined as squares in phase space of dimensions
$\left(\SI{0.15}{\delta},2\right)$,
representing the bottom-left and top-right corners of Fig.\,\ref{fig:transition_histogram}.
Paths that connect separate basins of attraction are registered in the histogram, 
giving the number of points in each bin $\mathcal{N}_{\left(x,N_{R}\right)}$
and the total number of points $\mathcal{N}_{transit}$, which is given by the sum over all bins,
\begin{equation}
\mathcal{N}_{transit} = \sum_{\{(x,N_{R})\}}\mathcal{N}_{\left(x,N_{R}\right)}.
\label{eq:number_transition_points}
\end{equation}
By estimating the probability
\begin{equation}
\mathcal{P}\!\left(x,N_{R}\mid \mathrm{transit} \right) = \frac{\mathcal{N}_{\left(x,N_{R}\right)}}{\mathcal{N}_{transit}},
\label{eq:probability_given_transition}
\end{equation}
we show in Fig.\,\ref{fig:transition_histogram} 
that transition paths typically follow the diagonal parameterized by $\alpha$.
Furthermore, we can define the transition state 
as the region of maximum transition probability~\cite{sup_hummer_transition_2003}.
This probability is calculated using Bayes' theorem,
\begin{equation}
\mathcal{P}\!\left(\mathrm{transit} \mid x,N_{R}\right) = 
\frac{\mathcal{P}\!\left(x,N_{R}\mid \mathrm{transit} \right) \mathcal{P}\!\left(\mathrm{transit} \right)}
{\mathcal{P}\!\left(x,N_{R}\right)} =
\frac{\mathcal{N}_{\left(x,N_{R}\right)}}{\mathcal{N}_{max}\mathcal{P}\!\left(x,N_{R}\right)}.
\label{eq:transition_prob_given_location}
\end{equation}
Before the estimation, $\mathcal{N}_{max}$ is known and $\mathcal{P}\!\left(x,N_{R}\right)$ can be calculated 
exactly from the free energy given by Eq.\ref{eq:helmholtz_free_energy} in the main text.
To further enhance the estimation process, we make use of the invariance under reflections where both
$R\leftrightarrow L$ and $x\leftrightarrow\delta-x$.
Fig.\,\ref{fig:committer_histogram} shows that 
the transition state is approximately perpendicular to the reaction coordinate
in the regions where we have sufficient statistics.
This supports the view that $\alpha$ characterizes transitions well,
and that the free energy barrier should be calculated as a function of $\alpha$.

%======================================================================

\section{\label{sec:approximation_free_energy} Exponential approximation of free energy barrier}
We have an analytical expression for the partition sum 
as a function of the reaction coordinate $\alpha$ in Eq.\,\ref{eq:reaction_coordinate_si},
\begin{align}
\mathcal{Z}\!\left(\alpha\right) &= \sum_{N_{R}=0}^{N-1} \mathbb{1}\!\left(0 \leq 2\alpha-\frac{N_{R}}{N} \leq 1 \right) 
\binom{\ell-N_{R}}{N-N_{R}}\binom{\ell-N+N_{R}}{N_{R}} 
\nonumber \\*
& \quad \times
\exp\!\left[ - \frac{k \delta^{2} N}{2 k_{B}T} \left(4\left(\alpha-\frac{N_{R}}{N}\right)^{2}
+\frac{N_{R}}{N}\left(1-\frac{N_{R}}{N}\right) \right) \right].
\label{eq:partition_sum_alpha}
\end{align}
The indicator function $\mathbb{1}$ makes sure that the summation is only performed over those terms 
that represent a valid value of the position $x$, which should obey $0\leq x \leq \delta$. 
The free energy barrier peak is located at $\alpha=1/2$, while the valleys are located at $\alpha=0$ and $\alpha=1$.
Hence, the height of the barrier is given by
\begin{equation}
\Delta \mathcal{F}^{\ddagger} = -k_{B}T \log \! \left(\frac{\mathcal{Z}\!\left(\alpha=1/2\right)}
{\mathcal{Z}\!\left(\alpha = 0\right)} \right).
\label{eq:definition_free_energy_barrier_height}
\end{equation}
At $\alpha=1/2$ all terms of the sum contribute to Eq.\,\ref{eq:partition_sum_alpha},
while at  $\alpha=0$, only the $N_{R}=0$ term contributes. 
The latter observation leads to a simple expression for the partition sum in the valley,
\begin{equation}
\mathcal{Z}\!\left(\alpha = 0\right) = \binom{\ell}{N}.
\label{eq:partition_sum_alpha_0}
\end{equation}
Then, using that
\begin{equation}
\frac{N_{R}}{N}\left(1-\frac{N_{R}}{N}\right) = \frac{1}{4} - \left(\frac{N_{R}}{N}-\frac{1}{2}\right)^{2} 
= \frac{1}{4} - \frac{1}{N^{2}}\left(N_{R}-\frac{N}{2}\right)^{2},
\label{eq:rewrite_product}
\end{equation}
we get
\begin{align}
\frac{\mathcal{Z}\!\left(\alpha=1/2\right)}{\mathcal{Z}\!\left(\alpha = 0\right)} 
&= \sum_{N_{R}=0}^{N-1} \binom{\ell-N_{R}}{N-N_{R}}\binom{\ell-N+N_{R}}{N_{R}} / \binom{\ell}{N}
\nonumber \\*
& \quad \times
\exp\!\left[ - \frac{k \delta^{2} N}{2 k_{B}T} \left(\frac{3}{N^{2}}\left(N_{R}-\frac{N}{2}\right)^{2}
+\frac{1}{4} \right) \right].
\label{eq:rewrite_partition_sum_barrier}
\end{align}
This equation gives the exact value of the free energy barrier height 
through Eq.\,\ref{eq:definition_free_energy_barrier_height}.

Even though we have an exact solution for the barrier height, 
Eq.\,\ref{eq:rewrite_partition_sum_barrier} does not provide any understanding on how the barrier height
depends on the number of crosslinkers or on the microtubule overlap length.
To acquire a better comprehension of these dependencies, 
we require an analytical approximation for the free energy barrier height in terms of simple functions.
Here, we make such an approximation using three conditions.
First, we assume that the crosslinker density $N/\ell$ is small. 
Second, we recognize that the summand peaks at $N_{R}=N/2$, 
and that we capture the main contribution to the sum by Taylor expanding the function around this point.
Third, we assume that $N$ is large enough such that the summand does not change too strongly as a function of $N_{R}$.
Under that last condition, we can replace the sum by an integral over the real line.

To start, we first rewrite the product of binomial coefficients,
\begin{equation}
\binom{\ell-N_{R}}{N-N_{R}}\binom{\ell-N+N_{R}}{N_{R}} / \binom{\ell}{N} 
= \binom{N}{N_{R}} \frac{\left(\ell-N_{R}\right)!\left(\ell-N+N_{R}\right)!}{\ell!\left(\ell-N\right)!}.
\label{eq:rewrite_binomials}
\end{equation}
The first binomial coefficient captures the main contribution of changes in $N_{R}$, 
while the second factor approaches unity for very low densities.
Now, we will make Gaussian approximations for the binomial coefficients, 
following a standard approach presented e.g. by Milewski~\cite{sup_milewski_derivation_2007}. 
First, we reparameterize the equation using $N_{R} = N/2+M/2$, or $M=2N_{R}-N$, which allows us to expand around $M=0$.
Then, we use Stirling's Approximation on all factorials in the binomial, 
the first factor of Eq.\,\ref{eq:rewrite_binomials} becomes
\begin{align}
\binom{N}{N_{R}} &= \binom{N}{\frac{N}{2}+\frac{M}{2}} 
= \frac{N!}{\left(\frac{N}{2}+\frac{M}{2}\right)!\left(\frac{N}{2}-\frac{M}{2}\right)!}\nonumber \\*
&\approx \frac{N^{N+\frac{1}{2}}}{\sqrt{2\pi} \left(\frac{N}{2}+\frac{M}{2}\right)^{\frac{N}{2}+\frac{M}{2}+\frac{1}{2}}
	\left(\frac{N}{2}-\frac{M}{2}\right)^{\frac{N}{2}-\frac{M}{2}+\frac{1}{2}}} \nonumber \\*
&= \frac{1}{\sqrt{2\pi}}
\frac{N^{N+\frac{1}{2}}}{\left(\frac{N^{2}}{4}-\frac{M^{2}}{4}\right)^{\frac{N}{2}+\frac{1}{2}}}
\left(\frac{\frac{N}{2}-\frac{M}{2}}{\frac{N}{2}+\frac{M}{2}}\right)^{\frac{M}{2}} \nonumber \\*
&= \frac{1}{\sqrt{2\pi}} \frac{N^{N+\frac{1}{2}} 2^{N+1}
	N^{-N-1}}{\left(1-\frac{M^{2}}{N^{2}}\right)^{\frac{N}{2}+\frac{1}{2}}}
\left(\frac{1-\frac{M}{N}}{1+\frac{M}{N}}\right)^{\frac{M}{2}} \nonumber \\*
&\approx \sqrt{\frac{2}{\pi N}} 2^{N} \left(1-\frac{M^{2}}{N^{2}}\right)^{-\left(\frac{N}{2}+\frac{1}{2}\right)}
\left(1-2\frac{M}{N}+2\left(\frac{M}{N}\right)^{2}\right)^{\frac{M}{2}}.
\label{eq:first_part_binomial_approximation}
\end{align}
In the first line, all exponential terms from Stirling's Approximation cancel, 
and only one $\sqrt{2\pi}$ factor survives.
Then, after rearranging the result, we apply the geometric series and keep terms that are at most quadratic in $M/N$.
The equation can be further approximated after applying the logarithm,
\begin{align}
\log\binom{N}{N_{R}} &\approx \log\!\left(\sqrt{\frac{2}{\pi N}} 2^{N}\right)
-\left(\frac{N}{2}+\frac{1}{2}\right) \log\!\left(1-\frac{M^{2}}{N^{2}}\right) 
+\frac{M}{2} \log\!\left(1-2\frac{M}{N}+2\left(\frac{M}{N}\right)^{2}\right)\nonumber \\*
&\approx  \log\!\left(\sqrt{\frac{2}{\pi N}} 2^{N}\right) + \left(\frac{N}{2}+\frac{1}{2}\right)\frac{M^{2}}{N^{2}}
-\frac{M^{2}}{N} \nonumber \\*
&\approx  \log\!\left(\sqrt{\frac{2}{\pi N}} 2^{N}\right) -\frac{M^{2}}{2N}.
\label{eq:second_part_binomial_approximation}
\end{align}
Here we used the Taylor expansion of the logarithm in $M$ around $0$, 
and used the assumption that $M$ is smaller than $N$.
We can exponentiate the result back to the desired approximation,
\begin{equation}
\binom{N}{N_{R}} \approx \sqrt{\frac{2}{\pi N}} 2^{N} \mathrm{e}^{-\frac{M^{2}}{2N}} = \frac{2}{\sqrt{\pi N}} 2^{N}
\mathrm{e}^{-\frac{2}{N} \left(N_{R}-\frac{N}{2}\right)^{2}}.
\label{eq:binomial_approximation}
\end{equation}
This is the well known Gaussian approximation of the binomial distribution for $p=\frac{1}{2}$.
We use the same methods to make an approximation of the second factor in Eq.\,\ref{eq:rewrite_binomials},
\begin{align}
\frac{\left(\ell-N_{R}\right)!\left(\ell-N+N_{R}\right)!}{\ell!\left(\ell-N\right)!} 
&\approx \frac{\left(\ell-N_{R}\right)^{\ell-N_{R}+\frac{1}{2}} \left(\ell-N+N_{R}\right)^{\ell-N+N_{R}+\frac{1}{2}}}
{\ell^{\ell+\frac{1}{2}}\left(\ell-N\right)^{\ell-N+\frac{1}{2}}} \nonumber \\*
&= \left(\frac{\left(1-\frac{N_{R}}{\ell}\right)\left(1-\frac{N-N_{R}}{\ell}\right)}
{\left(1-\frac{N}{\ell}\right)}\right)^{\ell+\frac{1}{2}}
\left(\frac{1-\frac{N}{\ell}}{1-\frac{N_{R}}{\ell}}\right)^{N_{R}}
\left(\frac{1-\frac{N}{\ell}}{1-\frac{N-N_{R}}{\ell}}\right)^{N-N_{R}} \nonumber \\*
&= \left(\frac{\left(1-\frac{N}{2\ell}-\frac{M}{2\ell}\right)\left(1-\frac{N}{2\ell}+\frac{M}{2\ell}\right)}
{\left(1-\frac{N}{\ell}\right)}\right)^{\ell+\frac{1}{2}}
\left(\frac{1-\frac{N}{\ell}}{1-\frac{N}{2\ell}-\frac{M}{2\ell}}\right)^{\frac{N}{2}+\frac{M}{2}}
\left(\frac{1-\frac{N}{\ell}}{1-\frac{N}{2\ell}+\frac{M}{2\ell}}\right)^{\frac{N}{2}-\frac{M}{2}} \nonumber \\*
&= \left(\frac{\left(1-\frac{N}{2\ell}-\frac{M}{2\ell}\right)\left(1-\frac{N}{2\ell}+\frac{M}{2\ell}\right)}
{\left(1-\frac{N}{\ell}\right)}\right)^{\ell-\frac{N}{2}+\frac{1}{2}}
\left(1-\frac{N}{\ell}\right)^{\frac{N}{2}}
\left(\frac{1-\frac{N}{2\ell}+\frac{M}{2\ell}}{1-\frac{N}{2\ell}-\frac{M}{2\ell}}\right)^{\frac{M}{2}}.
\label{eq:first_part_factorial_approximation}
\end{align}
In the first line we applied Stirling's approximation again, 
after which we that both the numerator and denominator contain $\left(2\ell-N+1\right)$ factors.
This means we can divide $\ell$ out of them, 
after which we rearrange the result into three factors with different exponents.
There, we also use that $N=N_{R}+\left(N-N_{R}\right)$.
Then, in the third line, we substitute our definition of $M$, 
and finally regroup the results according to their new exponents.

We continue by approximating the first factor of Eq.\,\ref{eq:first_part_factorial_approximation}.
We apply the geometric series and only keep factors of quadratic order,
\begin{align}
\frac{\left(1-\frac{N}{2\ell}-\frac{M}{2\ell}\right)\left(1-\frac{N}{2\ell}+\frac{M}{2\ell}\right)}
{\left(1-\frac{N}{\ell}\right)}
&\approx \left(1-\frac{N}{\ell}+\frac{N^{2}}{4\ell^{2}}-\frac{M^{2}}{4\ell^{2}}\right)
\left(1+\frac{N}{\ell}+\frac{N^{2}}{\ell^{2}}\right) \nonumber \\*
&\approx 1+\frac{N^{2}}{4\ell^{2}}-\frac{M^{2}}{4\ell^{2}}.
\label{eq:second_part_factorial_approximation}
\end{align}
Similarly, we approximate the final factor of Eq.\,\ref{eq:first_part_factorial_approximation},
\begin{align}
\frac{1-\frac{N}{2\ell}+\frac{M}{2\ell}}{1-\frac{N}{2\ell}-\frac{M}{2\ell}} 
&\approx \left(1-\frac{N}{2\ell}+\frac{M}{2\ell}\right)
\left(1+\frac{N}{2\ell}+\frac{M}{2\ell}+\left(\frac{N}{2\ell}+\frac{M}{2\ell}\right)^{2}\right) \nonumber \\*
&\approx 1+ \frac{M}{\ell} +2 \left(\frac{M}{\ell}\right)^{2} + 
2 \left(\frac{N}{\ell}\right)\left(\frac{M}{\ell}\right).
\label{eq:third_part_factorial_approximation}
\end{align}
Then, we can take the logarithm of Eq.\,\ref{eq:first_part_factorial_approximation} and expand the logarithms,
\begin{align}
\log\!\left(\frac{\left(\ell-N_{R}\right)!\left(\ell-N+N_{R}\right)!}{\ell!\left(\ell-N\right)!} \right)
&\approx \left(\ell-\frac{N}{2}+\frac{1}{2}\right) 
\log\!\left(1+\frac{N^{2}}{4\ell^{2}}-\frac{M^{2}}{4\ell^{2}}\right) \nonumber \\*
&+ \frac{N}{2} \log\!\left(1-\frac{N}{\ell}\right) \\*
&+ \frac{M}{2} \log\!\left(1+ \frac{M}{\ell} +2 \left(\frac{M}{\ell}\right)^{2} 
+ 2 \left(\frac{N}{\ell}\right)\left(\frac{M}{\ell}\right)\right) \nonumber \\*
&\approx \frac{N^{2}}{4\ell} - \frac{M^{2}}{4\ell} - \frac{N^{2}}{2\ell} + \frac{M^{2}}{2\ell} \nonumber \\*
&= -\frac{N^{2}}{4\ell} + \frac{M^{2}}{4\ell} = -\frac{N^{2}}{4\ell} + \frac{\left(N_{R}-\frac{N}{2}\right)^{2}}{\ell}.
\label{eq:factorial_approximation}
\end{align}
After the expansion of the logarithms, we only keep terms that are at most quadratic in $M$ and $N$,
and that do not decay faster than $1/\ell$.
By combining the results from Eq.\,\ref{eq:binomial_approximation} and Eq.\,\ref{eq:factorial_approximation}, we find
\begin{equation}
\binom{\ell-N_{R}}{N-N_{R}}\binom{\ell-N+N_{R}}{N_{R}} / \binom{\ell}{N} 
\approx \sqrt{\frac{2}{\pi N}} \exp\!\left[N\log\!\left(2\right)-\frac{N^{2}}{4\ell}
-\left(\frac{2}{N}-\frac{1}{\ell}\right) \left(N_{R}-\frac{N}{2}\right)^{2}\right].
\label{eq:gaussian_approximation_binomials}
\end{equation}
Now we have the right tools to make an approximation of 
the barrier probability Eq.\,\ref{eq:rewrite_partition_sum_barrier},
\begin{equation}
\frac{\mathcal{Z}\!\left(\alpha=1/2\right)}{\mathcal{Z}\!\left(\alpha = 0\right)} 
\approx \sqrt{\frac{2}{\pi N}}
\exp\!\left(N\log\!\left(2\right)-\frac{N^{2}}{4\ell}
- \frac{k \delta^{2} N}{8 k_{B}T}\right)
\sum_{N_{R}=0}^{N-1} \exp\!\left[-\left(\frac{3 k \delta^{2}}{2 N k_{B}T}+\frac{2}{N}-\frac{1}{\ell}\right)
\left(N_{R}-\frac{N}{2}\right)^{2}\right].
\label{eq:barrier_probability_approximation_1}
\end{equation}
We make the variable substitution $\nu = N_{R}-N/2$ and treat it as a continuous variable, 
changing the sum into an integral and extending the summation region to the entire real line,
\begin{align}
\frac{\mathcal{Z}\!\left(\alpha=1/2\right)}{\mathcal{Z}\!\left(\alpha = 0\right)} 
&\approx \sqrt{\frac{2}{\pi N}}
\exp\!\left(N\log\!\left(2\right)-\frac{N^{2}}{4\ell}
- \frac{k \delta^{2} N}{8 k_{B}T}\right)
\int_{-\infty}^{\infty} \! \diff{\nu} 
\exp\!\left[-\left(\frac{3 k \delta^{2}}{2 N k_{B}T}+\frac{2}{N}-\frac{1}{\ell}\right) \nu^{2}\right] \nonumber \\*
&= \sqrt{\frac{2 \pi}{\pi N \left(\frac{3 k \delta^{2}}{2 N k_{B}T}+\frac{2}{N}-\frac{1}{\ell}\right)}}
\exp\!\left(N\log\!\left(2\right)-\frac{N^{2}}{4\ell}
- \frac{k \delta^{2} N}{8 k_{B}T}\right) \nonumber \\*
&\approx \frac{1}{\sqrt{1+\frac{3 k \delta^{2}}{4 k_{B}T}}}
\exp\!\left(N\log\!\left(2\right)-\frac{N^{2}}{4\ell}
- \frac{k \delta^{2} N}{8 k_{B}T}\right).
\label{eq:barrier_probability_approximation_2}
\end{align}
\begin{figure}
	\includegraphics{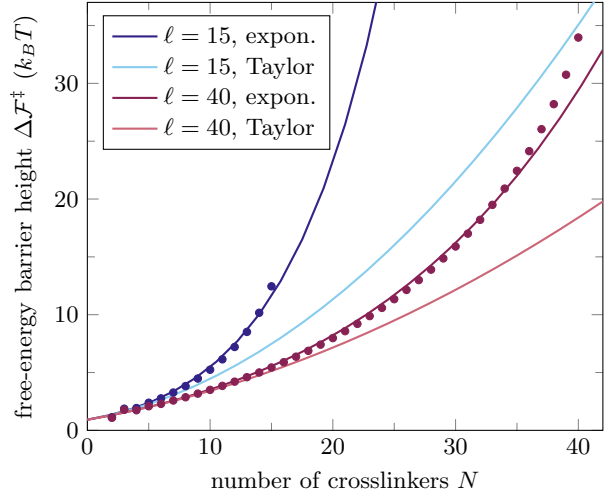}
	\caption{\label{fig:barrier_height_vs_n_approximations}
		The height of the free energy barrier $\Delta\mathcal{F}^{\ddagger}$ 
		as a function of the number of crosslinkers $N$ and two values of the mobile microtubule length $\ell$.
		The analytical solutions as calculated through 
		Eq.\,\ref{eq:definition_free_energy_barrier_height} and Eq.\,\ref{eq:rewrite_partition_sum_barrier}
		are plotted as sets of points.
		Lighter colors show the Taylor approximation Eq.\,\ref{eq:free_energy_barrier_taylor_approximation},
		and darker colors show the exponential approximation Eq.\,\ref{eq:free_energy_barrier_approximation}
		as well as the exact values.
		The Taylor approximations break down at relatively low crosslinker densities,
		but they predict the required exponent very well.
		The exponential approximations nearly perfectly follow the true barrier heights, 
		and only start to deviate at the highest crosslinker densities,
		where the true barrier height starts increasing even faster than exponentially.
	}
\end{figure}%
In the last line, we used that $k$ is relatively large, and that $N/\ell$ is small.
We find the approximation of the free energy barrier height using Eq.\,\ref{eq:definition_free_energy_barrier_height},
\begin{align}
\Delta \mathcal{F}^{\ddagger} &\approx \frac{k_{B}T}{2} \log \! \left(1+\frac{3 k \delta^{2}}{4 k_{B}T}\right)
+\left(\frac{k \delta^{2}}{8} - k_{B}T \log\!\left(2\right)\right)N + \frac{k_{B}T}{4\ell}N^{2} 
\label{eq:free_energy_barrier_taylor_approximation} \\*
&= \frac{k_{B}T}{2} \log \! \left(1+\frac{3 k \delta^{2}}{4 k_{B}T}\right)
+\left(\frac{k \delta^{2}}{8} - k_{B}T \log\!\left(2\right)\right)N\left(1+ \frac{1}{\frac{k \delta^{2}}{2k_{B}T} 
	- 4\log\!\left(2\right)}\frac{N}{\ell} \right) \nonumber \\*
&\approx \frac{k_{B}T}{2} \log \! \left(1+\frac{3 k \delta^{2}}{4 k_{B}T}\right)
+\left(\frac{k \delta^{2}}{8} - k_{B}T \log\!\left(2\right)\right)N\exp\!\left(\frac{1}{\frac{k \delta^{2}}{2k_{B}T} 
	- 4\log\!\left(2\right)}\frac{N}{\ell} \right).
\label{eq:free_energy_barrier_approximation}
\end{align}
The second line rewrites the result in terms of the number of crosslinkers $N$ and the crosslinker density $N/\ell$.
Then, in the final line, we exponentiate the part of the last term that depends on the density. 
This was done after inspection of the analytical free energy barrier height as a function of $N$,
of which examples are plotted as sets of points in Fig.\,\ref{fig:barrier_height_vs_n_approximations}.

The exponential version of the approximation captures the behavior of the exact function for a much larger range
than the second order Taylor approximation.
Furthermore, this version takes into account the variables that are intuitively important for the barrier height;
for small densities, the height increases linearly with the number of crosslinkers,
since a transition depends on the independent hopping of all crosslinkers.
Then, at high densities, exclusion effects start playing a role. 
Hence, we need to include a term that depends on the crosslinker density $N/\ell$,
which is done by the exponential term in Eq.\,\ref{eq:free_energy_barrier_approximation}.

%======================================================================


\end{document}